\documentclass[11pt, article]{amsart}
\usepackage{geometry}                
\usepackage{graphicx}
\usepackage{epstopdf}

\title{Landau hydrodynamic solution for non-central heavy ion collisions}
\author{Karolis Tamosiunas}                    
\address { Universidad Tecnica Federico Santa Maria \\ Casilla 110-V, Valparaiso \\ Chile }
\address { Institute of Theoretical Physics and Astronomy of Vilnius University \\ Go\v stauto g. 12 \\ 01108 Vilnius \\ Lithuania }
\email{karolis@itpa.lt}
\date{\today}
\begin{document}
\maketitle

\begin{abstract} 
This letter presents generalization of the Landau hydrodynamic solution for multiparticle production applied for non-central relativistic heavy ion collisions. Obtained solution enables to calculate anisotropy parameter $v_2$ for different rapidities with respect to the initial conditions for hydrodynamics. \\
PACS:  24.10.Nz - Hydrodynamic models, 25.75.Ag - Global features in relativistic heavy ion collisions, 25.75.Ld - Collective flow

\end{abstract}
%
%

\newcommand{\beq}{\begin{equation}}
\newcommand{\eeq}[1]{\label{#1} \end{equation}}
\newcommand{\bdm}{\begin{displaymath}}
\newcommand{\edm}{\end{displaymath}}

Landau's approximate hydrodynamic solution for particle production in relativistic heavy ion reactions \cite{Landau56} \cite{Landau53} is still valid for RHIC energies and reproduce particle spectra for different pseudorapidities well \cite{Steinberg:2004vy}.  
Experimentally observed azimuthal asymmetries of particle production in non-central heavy ion collisions are of high interest nowadays, as it does provide more information about the early dynamics of the high-energy nuclear reactions. Moreover, RHIC data \cite{rap_dist_rhic} of elliptic flow for different pseudorapidities shows universal behavior for different nuclei and for different beam energies, which is obtained from the model.  

The purpose of this letter is to introduce an analytic solution for relativistic hydrodynamic equations, including  azymuthally asymmetric transverse expansion.  
The presented solution is approximate, but in comparison to the computational hydrodynamics is analytic and transparent. The main assumptions of the model coinsides with original Landau approximations and are: 
i) longitudinal and transverse parts of hydrodynamic equations are solved separately; 
ii) equation of state of ideal relativistic gas, $p=e/3$, is used to solve transport equations;
iii) transverse expansion does not include initial flow and is pressure gradient driven. 

In this model we solve the equations of energy-momentum conservation:
\beq
\partial_\mu T^{\mu\nu}=0 ,
\eeq{hydroT}
where the energy-momentum tensor reads as: 
\beq
T^{\mu\nu}=(e+p)u^\mu u^\nu -pg^{\mu \nu}.
\eeq{def_T4}
The solution of local conservation laws (eq. \ref{hydroT}) with the equation of state represents the dynamics of the system by relating bulk properties of the matter, such as: energy density, $e$, pressure, $p$ and the four-velocity of the fluid, $u^\mu=u^0(1, \vec v)$.

The equations of the hydrodynamic longitudinal expansion in 1+1 dimension, along $z$ axis reads as: 
\beq 
\frac {\partial T^{00}}{\partial t} + \frac{\partial T^{0z}}{\partial z}=0 , \  \
\frac {\partial T^{0z}}{\partial t} + \frac{\partial T^{zz}}{\partial z}=0.
\eeq{long2}
Solution of the above equations starts by transforming relativistic velocity field components to rapidity terms, as:
\beq
u^0=\cosh \eta , \  \   u^z=\sinh \eta .
\eeq{u_to_rap}
From the above transformation, naturally follows, that: $v_z=\tanh \eta$ and $(u^z)^2 - (u^0)^2=-1$. 
The details on how to solve the equations of hydrodynamics (\ref{long2}) can be found in \cite{Landau56}, \cite{Landau53}, \cite{Khalatnikov54}, \cite{Khalatnikov04}, \cite{Wong:2008ex} so here we present only the reslt, as we will need it later.  
For abbreviation following variables are used: 
\begin{eqnarray}
\eta_+=1/2 \ln((t+z)/\Delta), \\
\eta_-=1/2 \ln((t-z)/\Delta),  
\end{eqnarray}
while $\Delta$ is the initial thickness of the system in the beam direction, $z$. Also, $\Delta$ is the initial condition after which equation of state assumed to be valid and evolution equations (\ref{long2}) are applied.

The final solution for energy density, $e(\eta_+,\eta_-)$, and rapidity, $\eta(\eta_+,\eta_-)$, is: 
\beq
e(\eta_+,\eta_-)=e_0 \exp[-4/3(\eta_+ + \eta_- -\sqrt {\eta_+ \eta_-})] ,
\eeq{en_dnz}
\beq
\eta(\eta_+,\eta_-)= ( \eta_+ - \eta_-)/2 , 
\eeq{z_y}
while $z=t\tanh \eta$. 
The above solution of (\ref{long2}) is exact and analytic of  1+1-dimensional relativistic hydrodynamics, which will be used after solving transverse expansion, to obtain the multiplicities of produced particles for different rapidities. 

In order to solve the transverse part of hydrodynamic equations we will follow original Landau assumptions with some modifications. For simplicity polar coordinates will be used, where four-flow in polar coordinates is: $
  u_i={dx_i}/{dt}$, $ u_0=(1-(\dot r^2 + r^2 \dot \phi^2))^{-1/2}, \ \ u_r=u_0v_r$ and energy-momentum tensor (\ref{def_T4}) components are as follows:  
 $$
T^{rr}= (e+p) (u^0)^2v_r^2+p, \ \  T^{\phi\phi}=(e+p)(u^0)^2v_\phi^2 + p/r^2
$$
$$
T^{0r}=(e+p) (u^0)^2v_r, \ \ T^{0\phi}=(e+p) (u^0)^2v_\phi
$$
Assuming that system does not rotate, $\dot\phi=v_\phi=0$, hydrodynamic equation (\ref{hydroT}) for the transverse dynamics at  the fixed transverse angle $\phi$ is: 
  \beq
\frac {\partial T^{0r}}{\partial t}  + \frac {\partial T^{rr} }{\partial r}=0
\eeq{tr_p}
Inserting energy-momentum tensor expressions to the above equation, we get: 
\beq
4e(u^0)^2\frac {\partial v_r}{\partial t} +4e(u^0)^2 \frac {\partial v_r^2 }{\partial r} +\frac {\partial e }{\partial r}=0
\eeq{tra1}
Following original Landau derivation, the fist term in the above equation is acceleration and assumed to be proportional to ${\partial v_r}/{\partial t}\propto r(t)/t^2$, the second term is set to zero, because  $v_r$ being comparably small. To simplify the third term Landau used $\partial e /\partial r \approx - e/R$, because the energy density at the center has value $e$ and is zero at the edge of the system, $r=R$. In the case of peripheral collisions, we do not expect centrally symmetric energy density distribution, thus the assumption is modified, as: 
\beq
\frac{\partial e}{\partial r}=\frac{e(r=R(\phi))-e(r=0)}{R(\phi)} ,
\eeq{tra2}
where $R(\phi)$ is a transverse radius of the system, which changes with the angle, as the system is not centrally symmetric. We do not know the value of $e(R(\phi))$, so we introduce a new function, as $f(R(\phi))=e(r=R(\phi))/e(r=0)$, which is a fraction of energy density at the edge of the system with respect to the energy density at the center.  
The function, $f(R(\phi))$, naturally must be less than unity for any $\phi$, as the energy density at the center is higher, than at the edge. Now from eq. (\ref{tra1}) we express transverse displacement dependance on time, as:  
\beq
r(t)=\frac{(1-f(R,\phi))t^2}{4(u^0)^2 R(\phi)} .
\eeq{rad}
The last stage of the model is so called conic-flight stage, where energy and entropy fluxes stop changing for the fixed cone element $2 \pi r dr$. With the help of the formula (\ref{rad}) we obtain a hypersurface in space-time, after which hydrodynamics stops and matter streams freely towards detectors. Here again we follow original Landau model, assuming a fixed transverse distance, $r(t)=a$, for the conic-flight to start. Thus, we obtain the value for the time, when the conic-flight starts, which reads as:  
\beq
t_1=2\cosh \eta \sqrt{\frac{aR(\phi)}{(1-f(R,\phi))}},  
\eeq{time}
where $u^0=\cosh \eta$. 

Solution in the conic-flight stage is straightforward, as the energy and entropy does not change at a fixed cone element. Connection of transverse and longitudinal solutions is done by matching the time $t=t_1$. Knowing, that  $dS=\sigma u^0 dz$ at a given time within element $dz$ and entropy density, $\sigma=ce^{3/4}$, we express entropy change over rapidity from the energy density formula (\ref{en_dnz}), as:  
\beq
\frac{dS}{d\eta}=ce_0^{3/4} \exp[-(\eta_+ + \eta_- -\sqrt {\eta_+ \eta_-})]\frac{t}{\cosh\eta}. 
\eeq{entr}
Inserting the solution for time eq. (\ref{time}) to the entropy equation above and assuming, that number of produced particles is directly proportional to the entropy,  $dN \propto dS$, one can obtain number of particles for different rapidities for a fixed angle $\phi$. What is still needed, is the function $f(R(\phi))$, hydrodynamic evolution length, $a$,  in  the transverse direction and initial starting thickness, $\Delta$. 
Besides, one has to normalize the particle distribution (\ref{entr}) with the total number of produced particles, to obtain the number of produced particles for different rapidities. But, from the definition of elliptic flow, $v_2$: 
\beq
v_2(\eta)=\frac{\int d\phi (dN/d\phi d\eta) \cos(2\phi)}{\int d\phi (dN/d\phi d\eta) }, 
\eeq{v2}
normalization of the total number of particles drops out, yet the dependance on the impact parameter $b$ is present. 
 
 
 In order to show that obtained hydrodynamic solution provides reliable results, we build the most simple initial configuration, which is based on Woods-Saxon nuclear density \cite{woodssaxon} and overlap geometry. The only assumption is, that nuclear density is directly proportional to the energy density. As we do not need the absolute value of energy density, this assumption is reasonable enough to obtain {\it qualitative} results. 
 
 The thickness function $T_{AA}(b,r,\phi)$ of nucleus is obtained, using Wood-Saxon nuclear density function\cite{woodssaxon}: 
$$
n_A(r)=\frac{n_0}{1+\exp(\frac{r-R_A}{d})} ,
$$
where $R_A=1.12A^{1/3}-0.86A^{-1/3}$ [fm], $d=0.54$ [fm] and $n_0=0.17 fm^{-3}$. Integrating nuclear density  $n_A(r)$ over beam axis $z$ one gets it's thickness: 
$$
T_A(x,y)=\int  dz \ n_A(\sqrt{x^2+y^2+z^2}). 
$$
The overlap thickness function with the distance $b$ between their centers is $
T_{AA}(b,x,y)=T_A(x-b/2,y)+T_A(x+b/2,y)$. The transformation to polar coordinates is straightforward with $x=r\cos\phi$, $y=r\sin\phi$. 
Finally, the function, $f(R(\phi))$, which gives the ratio of energy density at the edge of the system with energy density at the center, with fixed $b$, reads as: 
\beq
f(R(\phi))=T_{AA}(b,R(\phi))/T_{AA}(b,0).  
\eeq{fnuor}
The radius of the system, $R(\phi)$, is obtained from the geometry on how two circles overlap, as: 
$R(\phi)^2 + R(\phi)b\cos{\phi}+\frac{b^2}{4}-R_A^2=0$. 

\begin{figure}
\resizebox{1.0\columnwidth}{!}{%
  \includegraphics{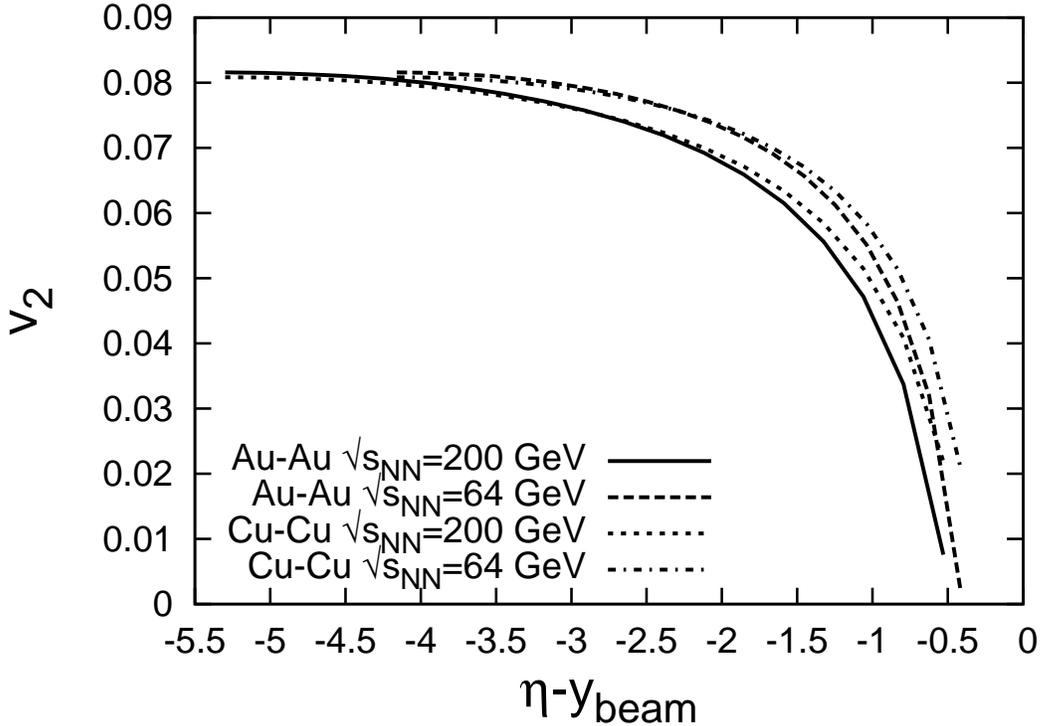}
}
\caption{Elliptic flow dependance on rapidity for Au-Au at $b=6$fm and Cu-Cu at $b=3$fm for two different collision energies. Beam rapidity $y_{beam}=\ln(\sqrt{s_{NN}}/m_N)$.}
\label{fig:1}       
\end{figure}

Merging equations (\ref{time}, \ref{entr}, \ref{fnuor}) we connect elliptic flow (\ref{v2}) with the initial thickness $\Delta$. Now it's possible to use experimentally observed elliptic flow for different pseudorapidities \cite{rap_dist_rhic}, which behaves universally for different nuclei collisions (Au-Au and Cu-Cu) and for different beam energies ($\sqrt{s_{NN}}=64$ and $200$  GeV). By comparing results with the data, we find, that hydrodynamic initial condition, $\Delta$,  must be proportional to the nuclear thickness, $\Delta=Cf(R(\phi))/\gamma$. The following means, that hydrodynamic expansion (\ref{long2}) starts with azimuthally asymmetric initial thickness, which is wider where initial  nuclear density is higher. Obtained results are shown in figure (\ref{fig:1}) for Au-Au at $b=6$fm and Cu-Cu at $b=3$fm reactions for two different beam energies, used at RHIC. We have used $C=1.5R_A$ and $a=R_A$ for all four curves in figure \ref{fig:1}. 

Quantitative fits to the initial thickness, $\Delta$, should be done with respect to more realistic picture of the energy density profile function, $f(R(\phi))$, which might be obtained from the distribution of participating nucleons and number of binary collisions with respect to the nuclear cross section, at the same time including beam energy dependance to the transverse dynamics. These more sophisticated models for the initial configuration are in development, but for now we can conclude, that Landau hydrodynamic solution and it's assumptions works not only for particle multiplicity spectra, but for elliptic flow and elliptic flow scaling as well. 

The generalized solution of Landau hydrodynamics incorporates transverse asymmetries into initial configuration and can be compared with wider amount of the experimental data. In this way giving deeper insight into the early dynamics of the system. The validity criteria of the obtained solution are no different from the original Landau hydrodynamic solution. 
 Using just a toy model, we have found that initial thickness of the system is not a flat disk, but is proportional to the azimuthally asymmetric nuclear density.

{\bf Acknowledgement} 
Author would like to acknowledge Centro Atomico of San Carlos de Bariliche, Argentina for their hospitality.


\end{document}